\documentclass[structabstract]{aa}
\usepackage{graphicx}
\usepackage{txfonts}
\begin{document}
   \title{Acoustic oscillations in stars near the tip of the red giant branch}
   \author{W. A. Dziembowski\inst{1,}\inst{2}
          \and
          I. Soszy\'nski\inst{1}}

\institute{Warsaw University Observatory,
              Aleje Ujazdowskie 4, 00-478 Warsaw, Poland\\
              \email {(wd,soszynsk)@astrouw.edu.pl}
         \and
             Copernicus Astronomical Center, ul. Bartycka 4, 00-787 Warsaw, Poland}
   \date{Received ??? ; accepted ???}


  \abstract
   {Small amplitude oscillations are observed in red giant branch(RGB) stars. Data on such
   oscillations are a source of information about the objects,
   notably about properties of convection in their envelopes and about the systems
   these objects inhabit. The OGLE-III catalog contains data for about 80 thousand
   small amplitude variable red giants (OSARGs) in the Large Magellanic Cloud.}
   {We want to explain variability in OSARGs as the solar-like oscillation and
   to associate the peaks in  power spectra with frequencies of acoustic modes.}
   {We use data on reddening-free magnitudes of the objects and interpret
   them in terms of stellar physical parameters using tabulated isochrones calculated for ages
   and composition parameters corresponding to the upper RGB of the LMC.
   Massive data on the peak frequencies and amplitudes
   are compared with expectations for stochastically excited oscillations.
   The frequencies  are also compared with those calculated for radial
   modes in envelope models with parameters taken from the isochrones.}
   {In stars close to the tip of the RGB, the peaks in power spectra are found in the
   0.1-1.0 $\mu$Hz range, which is consistent with extrapolation of the  frequency-luminosity relation
   for the solar-like oscillation. The dominant peaks occur close to  the first two radial overtones.
   The increase in amplitude with luminosity is slower than
   linear. The exponent $s\approx0.9$ is similar to what is found from recent analysis of CoRoT data on
   less luminous red giants. Frequency separations between dominant peaks are found to be smaller by about
   20 \% than calculated separations between these modes. After examining various possibilities,
   we left this discrepancy unexplained.}
   {The small amplitude variability of stars at the RGB tip is likely to be caused by a stochastic excitation
   of acoustic oscillations but interpreting of individual peaks in power spectra presents a problem.}


   \maketitle
%

   \begin{figure*}
   \centering
   \includegraphics{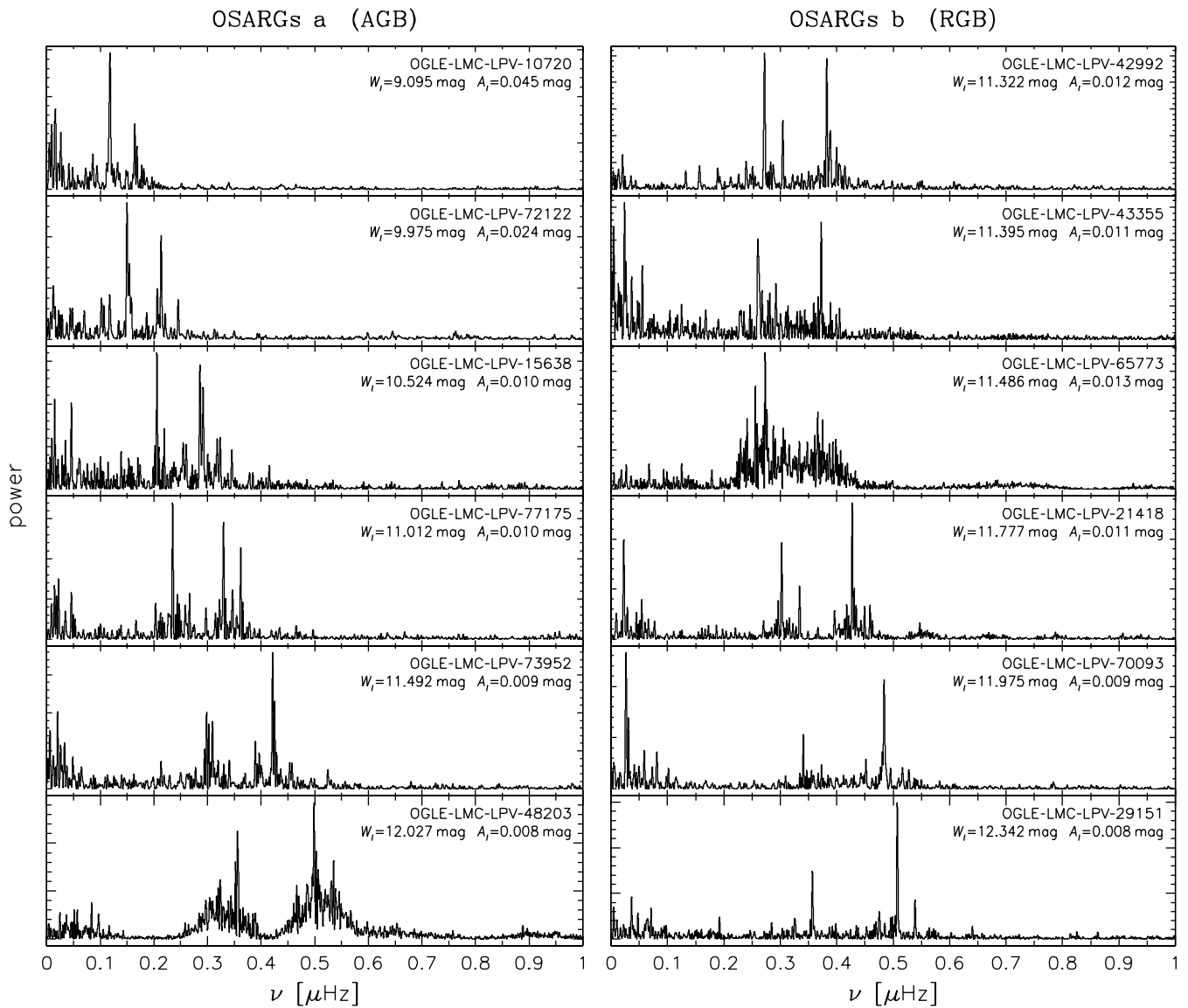}
   \caption{A sample of power spectra of a and b type OSARGs.
   For each spectrum we give the OGLE identification number, the
   Wesenheit index ($W_I$, see Eq. 1), and the amplitude of the highest peak in
   the the I-band ($A_I$).}
              \label{Fig1}%
    \end{figure*}

\section{Introduction}
It is now generally accepted that acoustic oscillations in the Sun
are stochastically excited by turbulent convection. Oscillations
driven in this way, called solar-like oscillations (SLOs), must be
excited in all stars having an outer convective zone but, owing to
small amplitudes, they are not easy to observe and study. The detection
of SLOs in stars along the red giant branch (RGB) was among the most
interesting results of asteroseismology with the space telescopes CoRoT
(De Ridder et al. 2009) and {\it KEPLER} (Stello et al 2010).
Oscillations of this type probe stellar convection,
so it is important to detect them in objects over a possibly wide
range of surface parameters. Data on mode frequencies in individual
red giants could give us important constraints on evolutionary
models. This requires, however, a credible mode identification.

The signature of SLOs is an excess of power around a given frequency
$\nu_{\rm max}$, which decreases with growing stellar luminosity.
The $L^{-1}$ decrease was expected (Kjeldsen \& Bedding 1995) and
partially confirmed by observations. Data from CoRoT (Mosser et al.
2010) and {\it KEPLER} (Stello et al. 2010) reach down to $\nu_{\rm
max}\approx2~\mu{\rm Hz}$, which roughly corresponds to
$L=300L_\odot$. This number is about one order of magnitude
above the {\it red clump} composed of helium-burning stars and one
order of magnitude below the RGB tip (TRGB), which stars reach at
the moment of helium ignition.

On a large scale, small amplitude oscillations in luminous RGB
and asymptotic giant branch (AGB) stars were detected as a byproduct
of microlensing surveys MACHO and OGLE. Using data
from the first survey, Wood et al. (1999) and Wood (2000) show
 that the LMC red giants form five ridges (sequences)
in the period-luminosity plane. The ones labeled A and B were
associated with the small amplitude pulsators. Later, the MACHO data
were analyzed by Fraser et al. (2005, 2008). Data from the OGLE-II survey
for both Magellanic Clouds were the subject of independent
analyses by Kiss \& Bedding (2003, 2004),  Soszy\'nski et al.
(2004),  and Ita et al. (2004a,b). Additional period-luminosity
sequences formed by low amplitude objects were identified by
these authors. Interpretations of these sequences in
terms of radial pulsation modes in RGB and AGB stars were considered.
Evidence for the multiperiodic character of pulsation was found in both
sets of data.

Here we use the acronym OSARGs {\it OGLE small amplitude red
giants}, which was introduced by Wray et al. (2004) in their paper
on variability of red giants in the Galactic bulge. Soszy\'nski et
al. (2004), who discovered a large number of OSARGs in the
Magellanic Clouds (15400 in LMC and 3000 in SMC), showed that they
fall into seven detached ridges in the period-luminosity plane. Four
of the ridges extending beyond the TRGB were attributed to the AGB
stars. These objects were named type a OSARGs. The ones lying within
the three ridges ending at TRGB were regarded genuine RGB stars and
named type b OSARGs.

The OGLE-III Catalog of Variable Stars (Soszy\'nski et al. 2009)
contains data for about $8\times10^4$ OSARGs in the Large Magellanic
Cloud (LMC) and is based on photometric data collected during 8
years (2001-2009) of continuous observations within the framework of
the OGLE-III project. For some of the objects, the time base was
increased to 13 years thanks to photometry from the OGLE-II project.
In this paper, we focus mainly on type b, because we only have
adequate models for RGB stars. First, in a phenomenological way, we
address the question how the OSARGs' variability is driven. Next, we
present an attempt to explain the dominant peaks in terms of stellar
oscillation modes.

\section{Are the OSARGs solar-like pulsators?}

The origin of OSARGs is not a new question. It has been already addressed by
Soszy\'nski et al. (2007), but some arguments in favor of such an
interpretation were presented in a relatively short section,
which went unnoticed by people working on SLOs.

Stochastic excitation must take place in red giants, but it is not at
all clear whether the observed variability of OSARGs is
predominantly caused by this mechanism. In fact, Xiong \& Deng
(2007) find that some low-order radial modes are unstable in RGB
stars. This instability arises mainly from a small excess of the
driving effect of turbulent pressure over damping effect of the
turbulent viscosity. Both effects originate in the perturbed
Raynolds stress, and are difficult to estimate. Thus both
the stochastic excitation and the self-excitation of unstable modes
must be regarded as viable suggestions. In the wake of uncertainties in
the description of convection in red giants, we propose a closer
look at observational data to see how the properties of low-amplitude oscillations
in luminous red giants follow the trends seen in fainter objects.

We began by comparing patterns in power spectra. Figure 1 shows
examples of spectra for OSARGs, based on the OGLE {\it I}-band
photometry. We chose the ones that have prominent peaks
contributing to a$_2$, a$_3$, b$_2$, b$_3$ sequences (Soszy\'nski et al. 2007),
which are the most prominent in the PL-plane. The patterns seen in this figure are
reminiscent of those shown in Fig. 1 of De Rider et al. (2009) or
Fig. 2 of Stello et al. (2010), except that the frequency range is a
factor 100 lower. In all periodograms shown in our Fig. 1, we see the
two prominent peaks that were the feature behind our selection rule,
but besides this, there is considerable diversity, especially in the
amount of power around the dominant peaks and in the presence of the
secondary peaks. In all cases there is a power localized in some
intermediate frequency range that moves leftwards with increasing
luminosity. Simultaneously, the separation between dominant peaks
decreases. These are features seen in all power spectra of SLOs.

The low-frequency power in the OGLE data may be attributed to the granulation noise, like in
other cool stars. However, the high-amplitude peaks in this range come from well-known but still
unexplained phenomenon of the {\it long secondary period} (LSP)
observed in a significant fraction ($\sim$30\%) of red giants (see,
e.g., Nicholls et al. 2009). The power spectra shown in this figure are very similar to those
shown by Tabur et al (2010) for bright Galactic M giants. Some of
those objects are evidently located near TRGB.

We turn now to data on all OSARGs that contribute to the sequences
b$_2$ and b$_3$. From the OGLE-III Catalog, we took the respective
frequencies and the Wesenheit index, $W_I$, for individual objects.
This index, which is the reddening-free stellar magnitude, is
defined as
\begin{equation}
W_I=I-1.55(V-I),
\end{equation}
where $V$ and $I$ denote mean magnitudes in the respective bands.
Points in Fig. 2 present data on the most significant periodicities
($S/N\ge5$). Up to two points are plotted per star. To compare
these data with information on SLOs in less luminous stars, we need
stellar parameters, such as luminosity, effective temperature, and
mass. To this end, we used isochrones downloaded from the BaSTI
Library (Pietrinferni et al. 2006) for ages and metal abundance
parameters appropriate to stars at TRGB in the LMC (Salaris \&
Girardi 2005). For the distance modulus to LMC, we adopted 18.5 mag,
which is close to the mean value from recent determinations by
various methods (Schaefer 2008). The characteristics of the
 selected isochrones and model parameters at the point where
 $W_I=12$~mag are listed in Table 1.
 In the range covered by the data,
 there  is about factor 4 increase in luminosity between the
 highest and the lowest frequencies.

\begin{table}
\caption{Selected isochrones and stellar parameters at $W_I=12$}
\label{table:1} \centering
\begin{tabular}{c c c c l l l }     
\hline
Age [Gy] & [M/H] & Z & Y & ${M/M_\odot}$ & ${L/L_\odot}$ & ${T_{\rm eff}/T_{{\rm eff},\odot}}$\\
\hline
   7 & -0.659 & 0.004 & 0.251 & 0.965 & 1795 & 0.657\\
   4 & -0.659 & 0.004 & 0.251 & 1.145 & 1945 & 0.658\\
   4 & -0.953 & 0.008 & 0.256 & 1.222 & 1563 & 0.642\\
\hline
\end{tabular}
\end{table}
%

   \begin{figure}
   \centering
   \resizebox{\hsize}{!}
   {\includegraphics{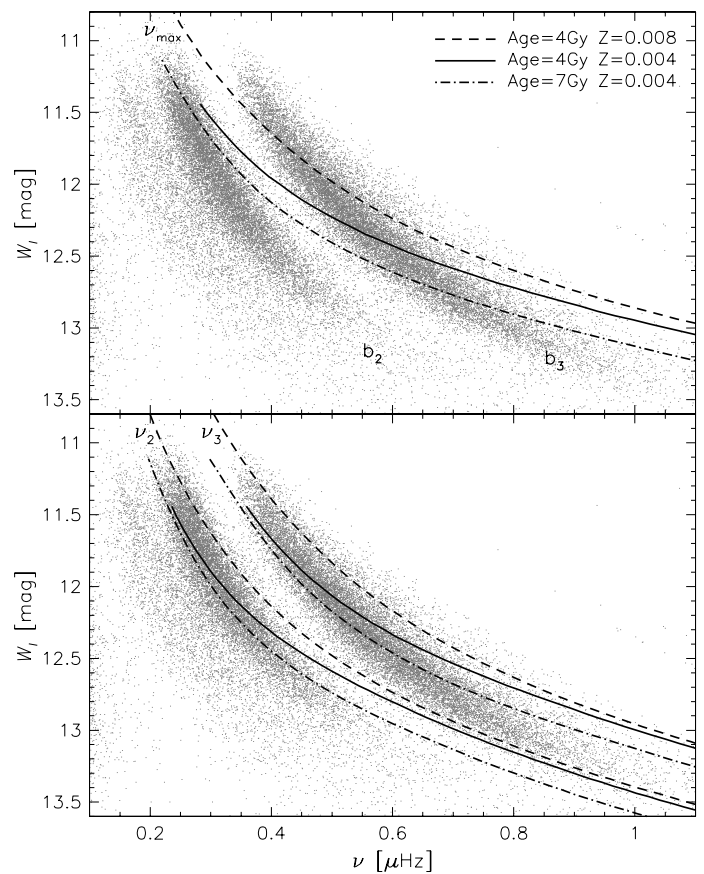}}
   \caption{The b$_1$ and b$_2$ sequences in the $\nu$-$W_I$ plane
   compared with the expected positions of power calculated with Eq.(2)
   ({\it top}) and with frequencies of
   the first two radial overtones, $\nu_2$ and $\nu_3$ ({\it bottom}). The
   dots show all significant peaks at the S/N=5 level. Some objects yield points
   on both sequences.}
   \label{Fig2}%
\end{figure}

   \begin{figure}
   \centering
   \resizebox{\hsize}{!}
   {\includegraphics{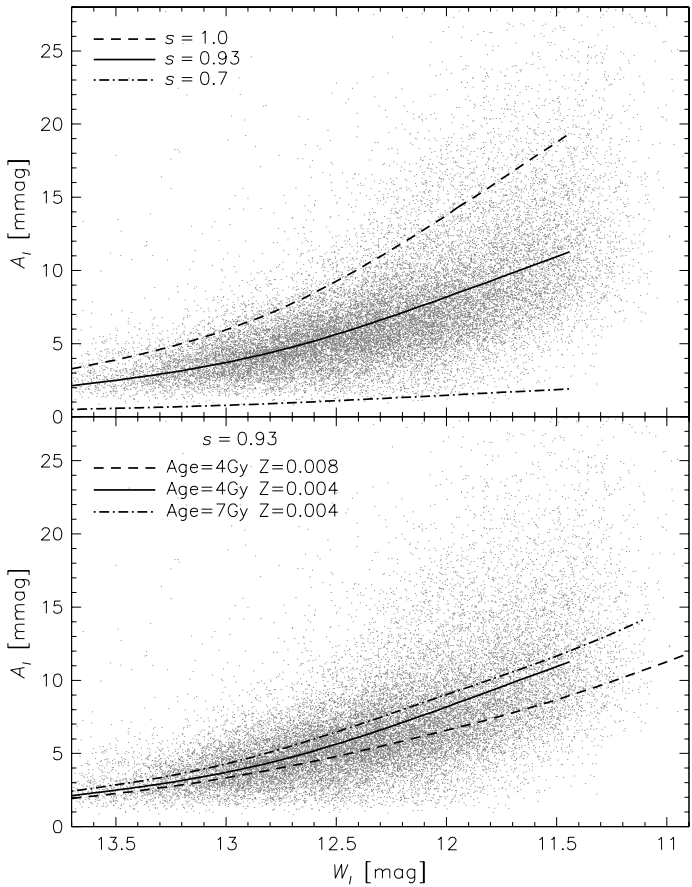}}
   \caption{The highet amplitudes of the peaks on the b$_2$ or b$_3$ sequences as a function
    of the reddening-free magnitude $W_I$ compared with the expected values for
    the solar-like oscillations calculated with Eq. (3). In the upper panel, the results of
    calculations with the three indicated values of $s$ for models along one of the
    three isochrones (Age=4Gy, $Z$=0.004) are shown. In the lower panel, results for the
    three isochrones at $s=0.93$ are compared }
   \label{Fig3}%
\end{figure}

The three lines in the upper panel of Fig. 2 were calculated along
the isochrones with the Kjeldsen \& Bedding (1995) expression
\begin{equation}
\nu_{\rm max}={L_\odot\over L}{M\over M_\odot}\left({T\over T_\odot}
\right)^{3.5}\times3050\mu Hz
\end{equation}
for the frequency corresponding to maximum of the acoustic power.
This expression is consistent with  data on solar-like oscillation
from main sequence up to lower RGB (Bedding \& Kjelsen 2003, Stello
et al. 2007). We may see that it also approximately applies to our
data, which extend up to TRGB. This is an argument in favor of the
solar-like nature of oscillations in type b OSARGs.  but the
possibility of self-excitation cannot be ruled out. The argument for
of the latter interpretation is that peaks are narrow, suggesting
long-lived modes.

In the lower panel we show the frequency-luminosity dependence for
the first and second overtones along the three isochrones.  The
frequencies were calculated with our nonadiabatic pulsation code
(Dziembowski, 1977) for envelope models with surface parameters from
the isochrones.\footnote{The parameters do not include the
surface helium abundance, which in the RGB stars may be  up to ten
percent higher than the initial one because of the dredge-up of the
nuclear processed gas. We checked that such an increase has no
visible effect on the plots shown in this paper.} The data suggest
that b$_2$ and b$_3$ sequences may be connected with excitation of
these two overtones. The frequency of the specified mode is roughly
proportional to the square of mean density, hence $\nu_k\propto
L^{-3/4}$. while $\nu_{\rm max}\propto L$. Therefore, with
increasing luminosity, the order of modes preferred by SLOs
decreases.

Another characteristic of the SLOs is the dependence of oscillation
amplitudes on stellar parameters. Kjeldsen \& Bedding (1995) derived
a semi-empirical expression for the bolometric luminosity amplitude
of $\delta L\propto(L/M)T_{\rm eff}^{-2}$. Samadi et al. (2007)
considered a more general dependence  of amplitude on the
luminosity-to-mass ratio, $(L/M)^s$. They show that the $s$-exponent
($sv$ in their notation) is determined by the eddy time-correlation
form. The Gaussian form yields $s=1$, while the Lorentzian $s=0.7$.
They also show that the latter value fits the ground-based data on
radial velocity amplitudes better. The space data on luminosity
amplitudes yield discrepant conclusions.  Stello et al. (2010) find
$s\approx0.7$ from the {\it Kepler} data, while Mosser et al. (2010)
found $s=0.89\pm0.02$ from their analysis of the CoRoT data.

The OGLE-III Catalog contains data on mode amplitudes in the $I$
band. We compared these data with the prediction from a
generalized form of Eq.(8) of Kjeldsen \& Bedding (1995), which at
$\lambda=800 {\mbox{ nm}}$ yields
\begin{equation}
A_I=\left({L\over L_\odot}{M_\odot\over
M}\right)^s\left({T_\odot\over T}\right)^2 \times 0.0035 {\rm mmag}.
\end{equation}
The plots in Fig. 3 show that the OGLE data follow the relation with
$s\approx0.9$. Both, $s=0.7$ and 1 are excluded. Our result is,
thus, similar to that of Mosser et al. (2010).

  \section{Modes excited in luminous red giants}
Identification of modes responsible for the peaks in power spectra
such as those shown in Fig. 1 would give us a basis for a real seismic
probing of individual luminous red giants. The plots in the bottom
panel of Fig.~2 suggest that the dominant peaks are associated with
the first and second overtones of radial pulsation. If it was true,
the frequency distance between the peaks would be the large
separation, $\Delta\nu$. In lower luminosity stars, more than one
large separation is measured. Their mean value, $<\!\Delta\nu\!>$,
is regarded as an important seismic parameter. However, the mean
distance between sequences b$_2$ and b$_3$ at specified $W_I$ cannot
be identified with $<\!\Delta\nu\!>$ because it does not refer to
the same object. We need the distance between the dominant peaks
measured separately in individual power spectra. With this in mind,
we selected over 100 objects with the most significant ($S/N\ge7$) peaks
within the two sequences. Plots in Fig. 4 show that, although the
observed $W_I(\nu_2)$ relation is reproduced well by the models, the
difference between the frequencies, which is shown in the lower
panel, is by some 20 percent less than the calculated difference
between the second and the first overtones. This difference is quite
a robust quantity depending almost only on $\nu_1$. It is only
weakly dependent on the isochrone parameters, as may be seen in the
figure, as well as on details of modeling the envelope and its
pulsation.

Our first suspicion has been that the discrepancy may be due to
inadequacies of our codes, which were written long ago (Paczy\'nski 1969; Dziembowski 1977).
The microphysics has been updated, but still the turbulent pressure and the Lagrangian
perturbation of the convective flux are ignored in these codes.
Fortunately, Xiong \& Deng (2004) have provided a functional expression for
radial mode frequencies calculated with their much more advanced
treatment of convection for a sequence of red giant envelope models.
Repeating calculation with our code for models with the same surface
parameters, we found only a small difference, which was much too
small to explain the discrepancy shown in Fig. 4. Also acceptable
departures from the mixing length parameter adopted in the
evolutionary models only have a minor effect on the dependence of
the $\nu_3-\nu_2$ difference on $\nu_2$.
   \begin{figure}
   \centering
   \resizebox{\hsize}{!}
 {\includegraphics{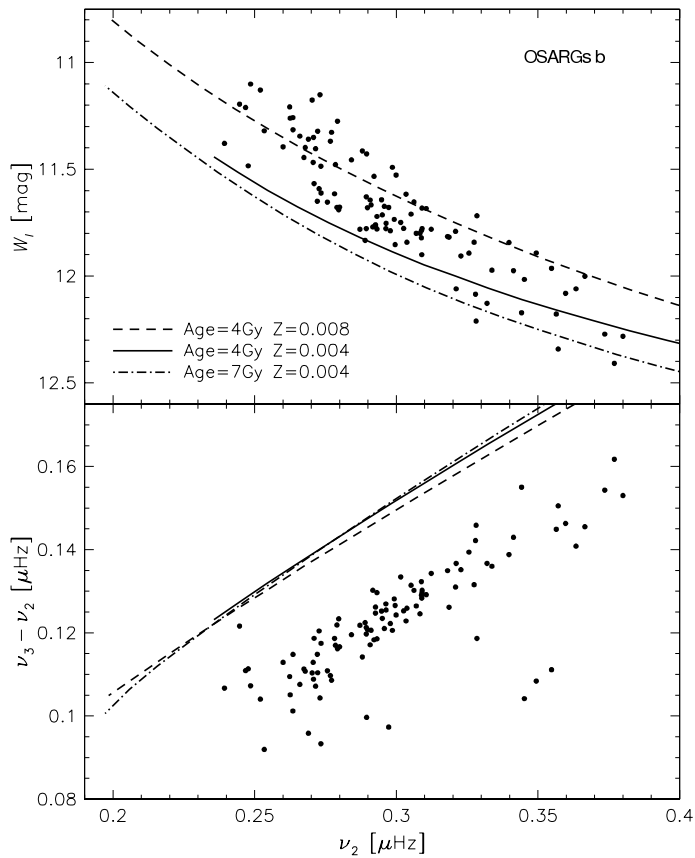}}
    \caption{Frequencies of the most significant peaks within the b$_2$ and b$_3$ sequences
    in selected objects (dots) compared with frequencies of the first two overtones in
    the models along the three isochrones. The top panel shows the reddening-free magnitude $W_I$
   as a function of the lower frequency. The bottom one shows the frequency difference.}
  \label{Fig4}%
  \end{figure}
   \begin{figure}
   \centering
   \resizebox{\hsize}{!}
 {\includegraphics{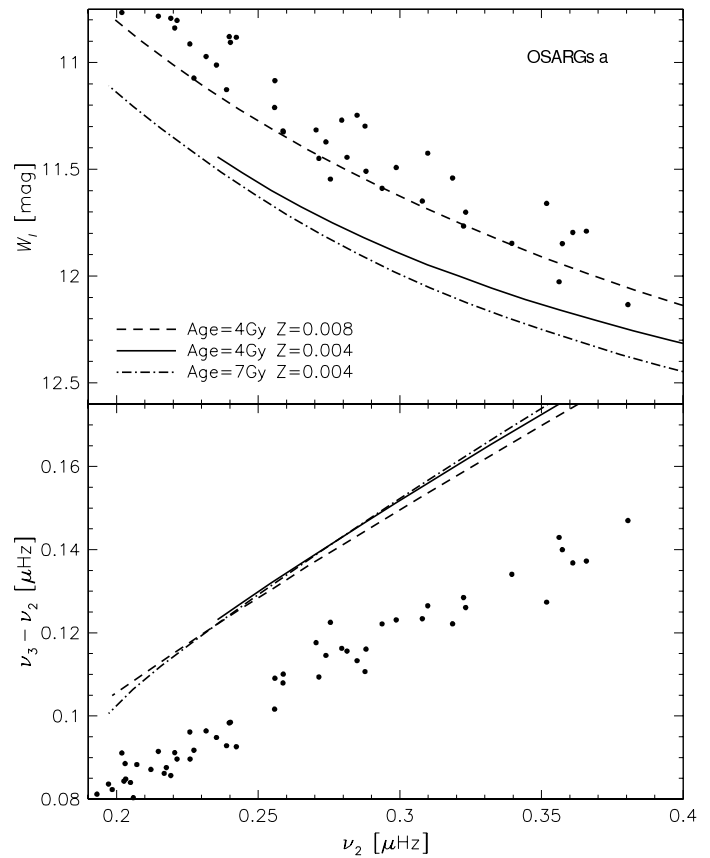}}
    \caption{Same as Fig. 4 but for peaks within the a$_2$ and a$_3$ sequences.}
  \label{Fig5}%
  \end{figure}

The problem with interpreting the b$_2$ and b$_3$ sequences in
terms of the first two overtones could have been noticed earlier
with a closer look at the Petersen Diagrams shown by Soszy\'nski et al.
(2004). In such diagrams, period ratios of all significant peaks
detected in individual objects are plotted against the longer
periods. In Fig. 11 of that paper, we may see a prominent cloud of
points corresponding to the b$_2$ and b$_3$ sequences near
$P_2/P_3\approx0.7$. However, most of them are located above this
value, whereas in the same period range, the models yield values
that are always below 0.7. The mean value for the cloud of 0.73 is
not reached until $\nu_2\approx1\mu$Hz.

We focused on type b OSARGs because they are, as we believe, RGB
objects, and we lack needed models of AGB stars. Nonetheless, it is
still instructive to compare frequencies of type a OSARGs with the
values calculated along the isochrones in the same range of the
$W_I$ magnitudes. The essential difference between AGB and RGB stars
concerns the deep interior, which has no effect
 on radial mode frequencies. The former objects have helium-exhausted cores, while the latter
are either in the phase of core helium burning or still ahead of it.
In Fig. 5, we compare data on type a OSARGs with results of the the
same calculations as were used for type b. Points in the upper panel
seem to follow a single isochrone, which should be somewhat younger
and/or have higher $Z$ than that for $Z=0.008$. This suggests that
at least some of type a OSARGs located below TRGB  are in fact the
first ascent objects before helium ignition. The data on frequency
differences shown in the lower panel reveal the same problem with
interpretation in terms of the frequency difference between the
first two radial overtones. The discrepancy is only somewhat smaller
but there is much less spread in the data. So the problem is not
less severe.

We regard the discrepancy revealed in Figs. 4 and 5 as serious. Any
other assignment of radial mode orders makes the discrepancy worse.
If no modification {in stellar models can remove the problem with
interpretation in terms of the radial modes, the option is to
consider non-radial modes trapped in the envelope. Such modes may
avoid large radiative losses in the interior (Dziembowski et al.
2001, Dupret et al 2009). Though it is difficult to explain why such
modes could be preferentially excited, we did consider this
possibility. We limited ourselves to dipolar modes because they
exhibit relatively small amplitude reduction caused by cancelation
of opposite sign contributions to disk-averaged value. The plots in
Fig. 6 show that allowing dipolar mode interpretation of the
dominant peaks does not solve the problem\footnote{The dipolar mode
frequencies were calculated with a new version of our code for
oscillations in unfitted envelope models that does not assume the
Cowling approximation.}. The vast majority of the data points shown
in the lower panel falls far from any of the three lines
representing calculated frequency differences.

 \begin{figure}
   \centering
   \resizebox{\hsize}{!}
 {\includegraphics{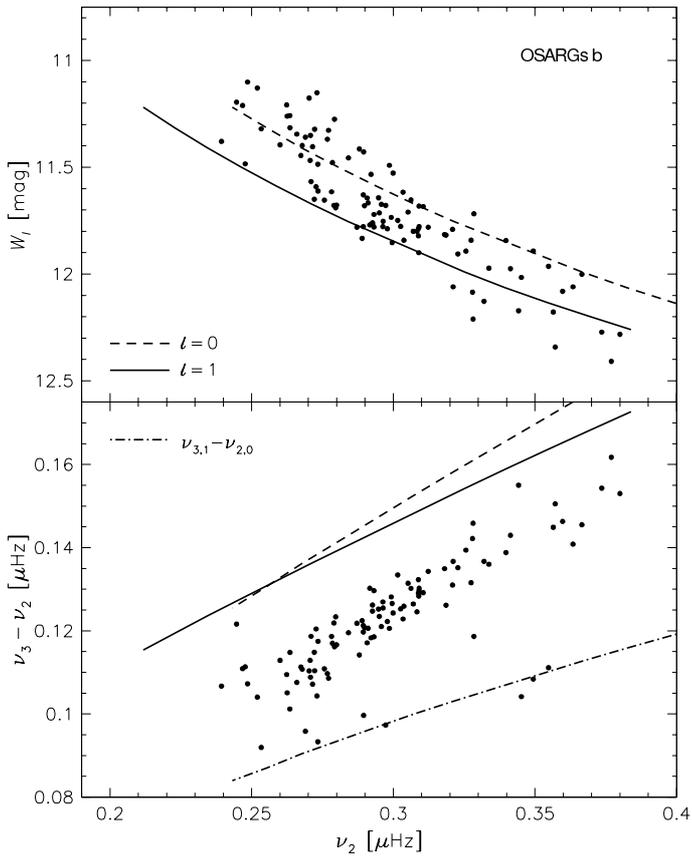}}
    \caption{Same data as in  Fig. 4  compared with frequencies of radial ($\ell=0$) and dipolar
    ($\ell=1$) modes in models along the Age=4Gy, Z=0.008 isochrone. The second subscript at $\nu$ corresponds to the
    angular degree, $\ell$}.
  \label{Fig5}%
  \end{figure}

\section{Discussion and conclusions}

Small-amplitude variability is a common phenomenon in all stars
along the RGB. Data from the ongoing space missions CoRoT and {\it
Kepler} filled the gap that existed between low-luminosity objects,
whose oscillations were found in radial velocity data, and ones
located near the tip detected by means of massive photometry. The
small-amplitude variability observed at lower luminosity is always
interpreted, as in the Sun, as a manifestation of a stochastic
excitation of acoustic modes.

The first question we addressed in this paper was whether
oscillations observed in the high-luminosity red giants from the
OGLE-III catalog (OSARGs) follow the pattern found at lower
luminosity. We showed that for these stars the oscillation power is
concentrated in the frequency range of the first two overtones,
which is consistent with extrapolation of the $\nu_{\rm max}(L)$
relation established for less luminous objects. For the solar-like
oscillations, amplitude increase with luminosity is described by the
$A\propto L^s$ relation, with $s$ in the [0.7-1.0] range. Samadi et
al.(2007) show that the value $s$ depends on properties of
turbulence, which drives the oscillations, and that data for
low-luminosity objects are best fitted with $s\approx0.7$. De Ridder
et al. (2009) obtained a similar result for more luminous stars. Our
value $s\approx0.9$ for the OGLE-III sample of RGB stars is similar
to the one found by Mosser et al. (2010) for fainter objects. These
arguments, in addition to the apparent similarity of power spectra
shown in  Fig. 1, support the idea that the variability of stars
near the tip of RGB is caused by a stochastic excitation of acoustic
modes.

We encountered a problem in our effort to assign radial mode
frequencies to the two dominant peaks in the OSARGs power spectra.
The observed frequency dependence on luminosity is not far from what
is expected for first two radial overtones, but the frequency
difference is by some 20\% less than that calculated for the models.
The calculated frequency difference was found to be only weakly
dependent on model parameters. We could not find any reasonable
modification in models that could lead us in the right direction. We
checked that the frequencies calculated with our code are in good
agreement with ones calculated for stellar envelope models with the
same surface parameters by Xiong \& Deng (2007). We considered the
possibility that one or both peaks  are stem from the dipolar mode
excitation and found that this does not solve the problem either.

Our failure does not mean the dominant peaks in OSARGs' power
spectra cannot be associated with low-degree modes of stellar
oscillations. The problem may still be caused by inadequacy of our
models. We do encourage further efforts along this line because
eigenmode frequencies are indeed the most important seismic
observables.

The alternative. which may seem less attractive, is that the peaks
are not caused by stellar normal mode excitation.  Most of narrow
peaks in red giants' power spectra from the space missions have not
been associated with stellar modes either. However, even if this is
so, the data on red giant variability remain of considerable
interest for stellar physics. They certainly carry information that
deserves to be extracted. In the case of OSARGs, the data demanding
interpretation are the seven ridges in the frequency-magnitude
plane.
\begin{acknowledgements}
      Part of this work was supported by the Polish MNiSW grant number N N203 379636.
\end{acknowledgements}


\begin{thebibliography}{}
\bibitem[]{} Bedding, T.~R., \& Kjeldsen, H.\ 2003, PASA, 20, 203
\bibitem[]{} De Ridder, J., Barban, D., Baudin, F.,
    et al. 2009, Nature, 459, 398
\bibitem[]{} Dupret, M. A., Belkacem, K., Samadi, R.,
  et al. 2009, A\&A, 506, 57
\bibitem[]{} Dziembowski, W. 1977, Acta Astron., 27, 95
\bibitem[]{} Dziembowski, W. A., Gough, D. O., Houdek, G,
\& Sienkiewicz, R. 2001, MNRAS, 328, 601
\bibitem[]{} Fraser, O. J., Hawley, S. L, Cook, K, H. \& Keller, S. C. 2005, AJ, 129, 768
\bibitem[]{} Fraser, O. J., Hawley, S. L, \& Cook, K, H.,
 2008, AJ, 136, 1242
\bibitem[]{} Ita, Y., Tanab\'e, T., Matsunaga, N., et al. 2004a, MNRAS, 347, 720
\bibitem[]{} Ita, Y., Tanab\'e, T., Matsunaga, N., et al. 2004a, MNRAS, 353, 705
\bibitem[]{} Kiss, L. L., \& Bedding, T. R. 2003,MNRAS, 343, L79
\bibitem[]{} Kiss, L. L., \& Bedding, T. R. 2004,MNRAS, 347, L83
\bibitem[]{} Kjeldsen, H. \& Bedding, T. R. 1995, A\&A, 293, 87
\bibitem[]{} Mosser, B., Belkacem, Goupil, M.-J. Goupil, et al.
2010, A\&A, 517, 22
\bibitem[]{} Nicholls, C. P., Wood, P. R., Cioni, \& Soszy\'nski, I. 2009, MNRAS, 399, 2063
\bibitem[]{} Paczy\'nski, B. 1969, Acta Astron., 19, 1
\bibitem[]{}Pietrinferni, A., Cassisi, S., Salaris, M., \& Castelli,
F. 2006, ApJ, 642, 797
\bibitem[]{} Salaris, M. \& Girardi, I.  2005, MNRAS, 357, 669
\bibitem[]{} Samadi, R., Georgobiani, D., Trempedach, R.,
    et al. 2007,  A\&A, 463, 297
\bibitem[]{} Schaefer, B.E. 2008, AJ, 135, 112
\bibitem[]{} Soszy\'nski, I., Udalski, A., Kubiak, M.,
    et al. 2004, Acta Astron., 54, 129
\bibitem[]{} Soszy\'nski, I., Dziembowski, W. A., Udalski, A.,
    et al. 2007,  Acta Astron., 57, 1
\bibitem[]{} Soszy\'nski, I., Udalski, A., Szyma\'nski, M.,
    et al. 2009, Acta Astron., 59, 239
\bibitem[]{} Stello, D., Bruntt, H., Kjeldsen, H.,  et al. 2007, MNRAS, 377, 584
\bibitem[]{} Stello, D., Basu, S., Bruntt, H. et al. 2010, ApJ, 713, 182
\bibitem[]{} Tabur, V., Bedding, T. R., Kiss, L. L., et al. 2010,, arXiv:1007.2974v1
\bibitem[]{} Wood, P. R. 2000, Publ. Astron. Soc. Aust., 17, 18
\bibitem[]{} Wood, P. R., Alcock, C., Allsman, R. A., et al. 1999, IAU Symp., 191, 151
\bibitem[]{} Wray, J. J., Eyer, L., \& Paczy\'nski, B. 2004, MNRAS, 349, 1059
\bibitem[]{} Xiong, D. R. \& Deng, I. 2007, MNRAS, 378, 1270
\end{thebibliography}
\end{document}